\documentclass[11pt,a4paper]{article}
\usepackage[utf8]{inputenc}
\usepackage{amsmath}
\usepackage{amsfonts}
\usepackage{amssymb}
\usepackage[left=2.cm,right=2.cm,top=2.5cm,bottom=2.5cm]{geometry}
\usepackage{slashed}
\usepackage{hyperref}
\usepackage{graphicx}
\usepackage{caption}
\usepackage[affil-sl,auth-sc]{authblk}

\usepackage{cite}

\usepackage{color}
\newcommand{\ie}{{\em i.e.} }
\newcommand{\eg}{{\em e.g.} }
\newcommand{\GeV}{{\rm GeV} }
\newcommand{\TeV}{{\rm TeV} }

\newcommand{\vev}[1]{\langle #1 \rangle}
\newcommand{\lrb}[1]{\left( #1 \right)}
\newcommand{\lrsb}[1]{\left[ #1 \right]}

\newcommand{\lrBigsb}[1]{\Big[ #1 \Big]}

\usepackage{ifthen}
\usepackage{tikz}

\newcommand{\eqs}[1]{\foreach\i[count=\NumArgs] in {#1}{}%
\ifthenelse{\equal{\NumArgs}{1}}{eq.~(\ref{#1})}%
{\ifthenelse{\equal{\NumArgs}{2}}%
{eqs.~\foreach\i[count=\q]in{#1}{\ifthenelse{\equal{\q}{\NumArgs}}{and (\ref{\i})}{(\ref{\i})~}}}%
{eqs.~\foreach\i[count=\q]in{#1}{\ifthenelse{\equal{\q}{\NumArgs}}{and (\ref{\i})}{(\ref{\i}),~}}}}}

\newcommand{\Eqs}[1]{\foreach\i[count=\NumArgs] in {#1}{}%
\ifthenelse{\equal{\NumArgs}{1}}{eq.~(\ref{#1})}%
{\ifthenelse{\equal{\NumArgs}{2}}%
{Eqs.~\foreach\i[count=\q]in{#1}{\ifthenelse{\equal{\q}{\NumArgs}}{and (\ref{\i})}{(\ref{\i})~}}}%
{Eqs.~\foreach\i[count=\q]in{#1}{\ifthenelse{\equal{\q}{\NumArgs}}{and (\ref{\i})}{(\ref{\i}),~}}}}}

\newcommand{\refs}[1]{\foreach\i[count=\NumArgs] in {#1}{}%
\ifthenelse{\equal{\NumArgs}{1}}{(\ref{#1})}%
{\ifthenelse{\equal{\NumArgs}{2}}%
{\foreach\i[count=\q]in{#1}{\ifthenelse{\equal{\q}{\NumArgs}}{and (\ref{\i})}{(\ref{\i})~}}}%
{\foreach\i[count=\q]in{#1}{\ifthenelse{\equal{\q}{\NumArgs}}{and (\ref{\i})}{(\ref{\i}),~}}}}}

\author{Dimitrios Karamitros\footnote{email: {\tt dimitrios.karamitros@ncbj.gov.pl}}}
\affil{\small National Centre for Nuclear Research, ul. Ho\.za 69, 00-681 Warsaw, Poland}

\title{Pseudo Nambu-Goldstone Dark Matter: Examples of Vanishing Direct Detection Cross Section}
\begin{document}

\maketitle

\begin{abstract} 
We consider cases where the dark matter-nucleon interaction is naturally suppressed.
We explicitly show that by extending the standard model scalar sector by a number of singlets, can lead to a vanishing 
direct detection cross section, if some softly broken symmetries are imposed in the dark sector. In particular, it is shown that 
if said symmetries are $SU(2)$ ($SU(N)$) and $U(1) \times S_N$, then the resulting pseudo-Nambu-Goldstone bosons can constitute
the dark matter of the Universe, while naturally explaining the missing signal in nuclear recoil experiments.
\end{abstract}

\section{Introduction}\label{sec:intro}
\setcounter{equation}{0}

The current status of direct detection experiments reduces the allowed number of dark matter (DM) models with DM particle masses around the electroweak (EW) scale 
(typically $\mathcal{O}(\GeV)-\mathcal{O}(\TeV)$), as indicated by recent results from the \textit{XENON1T} collaboration~\cite{Aprile:2018dbl}. The main reason for 
this  is the incompatibility of the experimental results with what one would expect from dimensional arguments (\ie the so-called {\em WIMP miracle}~\cite{Bertone:2016nfn}),
indicating that a DM particle with mass around the EW scale, should have interactions with an EW strength. That is, if the DM freezes-out due to its annihilation
to standard model (SM) particles, its interaction  with nucleons should be of similar magnitude. 
Thus, if this assumption holds, the DM annihilation and its direct detection rate should be correlated, and nuclear recoil experiments (which have access to 
DM in the EW scale) should have already detected DM.

There are various ways that the missing direct detection signal can be explained. Interesting possibilities include those that do not follow 
standard DM annihilation to SM particles and alter the way the DM freezes-out,~\footnote{By doing this, the DM annihilation and
nucleon interaction are not correlated, in contrast to what one would expect from the dimensional argument above.} 
for example ``secluded"~\cite{Pospelov:2007mp,Pospelov:2008jd}  and ``cannibal"~\cite{Pappadopulo:2016pkp} DM models.
Other possibilities include suppression  of the  interaction between DM and the nucleons due to a heavy (integrated-out) mediator~\cite{Mebane:2013zga,Fedderke:2014wda,Hisano:2014kua,Matsumoto:2016hbs,Dedes:2016odh,Yepes:2018zkk}, the appearance of  
``blind spots"~\cite{Cheung:2012qy,Banerjee:2016hsk,Han:2018gej} or the smallness of the DM mass~\cite{Heurtier:2016iac,Dedes:2017shn,Knapen:2017xzo,Darme:2017glc,Dutra:2018gmv,Foldenauer:2018zrz,Matsumoto:2018acr} (including  
in some cases ``frozen-in" DM~\cite{Hall:2009bx}) which
makes the DM particle inaccessible to such experiments.~\footnote{However, this could change soon as the effort for detection of light DM
intensifies~\cite{Dedes:2009bk,Essig:2017kqs,Evans:2017lvd,Feng:2017uoz,Ariga:2018uku,Agnes:2018oej,Akesson:2018vlm}.} 
Among particularly appealing scenarios, however, direct detection experiments are unable to detect the WIMP due to symmetry arguments~\cite{Dedes:2014hga,Arcadi:2016kmk,Gross:2017dan,Balkin:2018tma,Alanne:2018zjm}.
In such models there is a symmetry that is responsible for the suppression of the  DM-nucleon cross-section, usually through the cancellation of the tree-level DM-nucleon interaction.

In the present work, we explore models that belong to the family of the so-called ``Higgs portal" DM models (\eg \cite{Silveira:1985rk,McDonald:1993ex,Burgess:2000yq,
Kim:2006af,LopezHonorez:2012kv,Freitas:2015hsa,Karam:2015jta}). Although  many models of DM coupled directly to the Higgs respect direct detection 
constraints (\eg \cite{Arcadi:2016qoz,Casas:2017jjg,Lopez-Honorez:2017ora,Filimonova:2018qdc}), this kind of DM opens up other interesting possibilities.
Our focus here is the pseudo-Nambu-Goldstone boson (PNGB) DM scenario. The general idea behind how this can help to evade direct detection bounds comes
from the observation that Nambu-Goldstone bosons (NGB), which result from a spontaneous breaking of a  global symmetry, have derivative couplings with other particles, 
and so their interactions vanish at zero momentum. On the other hand, a PNGB (a DM cannot be NGB, since it should be massive) is a result of a spontaneously broken 
approximate global symmetry, which could  induce new interactions resulting to a non-vanishing direct detection cross section. However, there are examples of a cancellation 
that allows the tree-level DM-nucleon interaction to vanish at the zero momentum transfer~\cite{Gross:2017dan,Alanne:2018zjm}, making models featuring such cancellation suitable 
DM candidates.

In our effort to identify PNGB models featuring the aforementioned cancellation, we extent the SM by a scalar field (singlet under the SM gauge symmetry) 
and doublet under a softly broken $SU(2)$ global symmetry.~\footnote{ Similar models have been studied in great detail~\cite{Bhattacharya:2013kma}, however we 
focus on the cancellation of the DM-nucleon cross section and show explicitly that this takes place regardless of the form of the soft breaking terms.}
We also show that the PNGBs in this case remain stable due to the symmetry properties of the interaction terms. 
Furthermore, we show how these arguments apply to a softly broken $SU(N)$ global symmetry.

Then, we move to another case,  where we add two scalar fields (again singlet under the SM), and we note that the cancellation of PNGB-nucleon interaction occurs assuming 
a permutation symmetry. However, in contrast to the minimal case~\cite{Gross:2017dan}, the PNGB is not naturally stable unless a dark CP--symmetry is imposed. We also show that this model can be generalized to an arbitrary number ($N$) of scalar fields, provided an $S_{N}$ symmetry assumption.

The outline of the paper is the following: in section~\ref{sec:SU(2)}, we discuss the DM content and the natural suppression of the DM-nucleon cross section in the 
$SU(2)$. At the end of this section, we also show how these results are generalized in the $SU(N)$ case. 
In sec.\ref{sec:S_N}, we consider the $U(1) \times S_2$ case, and show how the cancellation of the direct detection cross section takes place, 
which we then generalize to $U(1) \times S_N$.
Finally, in section~\ref{sec:Conclusion} we summarize our results, and comment on possible future directions.

\section{The $SU(2)$ case}\label{sec:SU(2)}
\setcounter{equation}{0}
In this section examine a dark sector with a softly broken $SU(2)$ symmetry, in order to determine if the cancellation takes place.   
Specifically, the SM is extended by a scalar ($\Phi$) which is a gauge singlet under the SM gauge group, and a doublet under a softly
broken $SU(2)$.  We show that indeed this model can provide us with naturally stable (multi-component) DM, which exhibits 
a cancellation of the DM-nucleon interaction. We also show that this holds for $SU(N)$ and $\Phi$ in the fundamental representation.~\footnote{There is {\tt python} module available (\href{https://github.com/dkaramit/pseudo-Goldstone_DM}{github.com/dkaramit/pseudo-Goldstone\_DM}) that can be used to obtain Feynman rules and {\tt LanHEP}~\cite{Semenov:2014rea} 
input files for the $SU(N)$ case.}

\subsubsection*{The potential and mass terms}
The potential is comprised of two parts, the symmetric and the soft breaking ones. The symmetric part (global $SU(2)$ invariant) is
\begin{equation}
V_{0}=-\frac{\mu_{H}^2}{2} |H|^2 + \frac{\lambda_{H}}{2}|H|^4  +
\frac{\lambda_{\Phi}}{2} |\Phi|^4 +\lambda_{H\Phi} |H|^2|\Phi|^2 \;,
\label{eq:V0_SU(2)}
\end{equation}
while the softly breaking part of the potential can be written as 
\begin{equation}
V_{\rm soft}= \sum\limits_{i=1}^{2} \sum\limits_{j=1}^{2} \lrsb {(m_{\Phi \, ij}^2 \Phi_{i} \Phi_{j}+{\rm h.c.}) + m_{\Phi \, ij}^{\prime \, 2} \Phi^{\dagger}_{i} \Phi_{j}}  \;,
\label{eq:Vsoft_SU(2)}
\end{equation}
with  $m_{\Phi \, 12}^2=m_{\Phi \, 21}^2$, $m_{\Phi \, 12}^{\prime \, 2}= (m_{\Phi \, 21}^{\prime \, 2})^{*}$, and $m_{\Phi \, 11,22}^{\prime \,2} \in \mathbb{R}$.  Also, note that the potential, 
$V=V_0 + V_{\rm soft}$, becomes $SU(2)$-invariant if  $m_{\Phi \, ij}=m_{\Phi \, 12}^{\prime}=0$ and $m_{\Phi \, 11}^{\prime \, 2}=m_{\Phi \, 22}^{\prime \, 2}$.
Assuming that both $H$ and $\Phi$ develop VEVs,
\begin{equation}
H= \frac{1}{\sqrt{2}} \left( \begin{matrix} 0 \\ h+ v  \end{matrix} \right), \;
\Phi= \frac{1}{\sqrt{2}} \left( \begin{matrix} \phi+ i s \\ \rho +i \chi + v_{\Phi}  \end{matrix} \right) \;,
\label{eq:vevs_SU2}
\end{equation}
where, without loss of generality we have  assumed that the lower component of $\Phi$ obtains a VEV.~\footnote{This can be done by a unitary transformation ($U_{X}$)
of some field $X$, with $\vev{X}=(v_{X_1},v_{X_2})$, to $\Phi$. In this case $v_{\Phi}=\sqrt{v_{X_1}^{2}+v_{X_2}^{2}}$, and $U_{X}=U_{X}(v_{X_{1,2}})$.\label{foot:Phi_rotation}}
By minimizing the potential, we obtain the following relations 
\begin{align}
&m_{\Phi \, 22}^2  \in  \mathbb{R} \nonumber \\
&\mu_{H}^2 =  \lambda_{H} v^2 + \lambda_{H\Phi} v_{\Phi}^2 \nonumber \\ 
&m_{\Phi \, 12}^{2} = - \frac{m_{\Phi \, 21}^{\prime \, 2}}{2} \nonumber \\
&\lambda_{\Phi} =  -\frac{1}{v_{\Phi}^2}\left[\lambda_{H \Phi}v^2 + 4 m_{\Phi_{22}}^2 + 2 m_{\Phi_{22}}^{\prime \, 2}  \right] \;,
\label{eq:extrema-SU(2)}
\end{align} 
where the first restriction is not an extra requirement as it is implied by the last relation.~\footnote{Note here that without any rotation
(see footnote~\ref{foot:Phi_rotation})~\eqs{eq:extrema-SU(2)} would relate the original VEVs to the parameters of the model. So, the relation
between $m_{\Phi \, 12}^{2}$ and $m_{\Phi \, 21}$ is, in fact, a relation between $v_{X_{1,2}}$ and other parameters of the model in the 
$X$ basis.}
The Lagrangian mass terms can be written as
\begin{equation}
\mathcal{L}_{\rm mass}=-\frac{1}{2} \Bigg(  G^{ T} M_{G}^2 G +  S^T M_{S}^2  S \Bigg)\;,
\end{equation}
where $G=(\chi, s, \phi)^T$ are the PNGBs and $S=(h, \rho)^T$. The mass matrices become 
\begin{align}
M_{G}^{2}=&\lrb{\begin{matrix}
	- 4 m_{\Phi_{22}}^{ 2 } & 2 \Re{\left(m_{\Phi_{12}}^{\prime \, 2 }\right)} &  - 2 \Im{m_{\Phi_{12}}^{\prime \, 2 }}
	\\[0.2cm]
	2 \Re{\left(m_{\Phi_{12}}^{\prime \, 2 }\right)} & - 2 m_{\Phi_{22}}^{ 2 } + m_{\Phi_{11}}^{\prime \, 2 } - m_{\Phi_{22}}^{\prime \, 2 } - 2 \Re{\left(m_{\Phi_{11}}^{ 2 }\right)} & - 2 \Im{m_{\Phi_{11}}^{ 2 }}
	\\[0.2cm]
	- 2 \Im{m_{\Phi_{12}}^{\prime \, 2 }} &- 2 \Im{m_{\Phi_{11}}^{ 2 }} &  - 2 m_{\Phi_{22}}^{ 2 } + m_{\Phi_{11}}^{\prime \, 2 } - m_{\Phi_{22}}^{\prime \, 2 } + 2 \Re{\left(m_{\Phi_{11}}^{ 2 }\right)}
	\end{matrix}} \nonumber   \\ 
& \nonumber   \\
M_S^2 =&\lrb{\begin{matrix} \lambda_H v^{2} &  \lambda_{H \Phi} v  v_{\Phi}\\ \lambda_{H \Phi} v  v_{\Phi} &   \lambda_{ \Phi} v_{\Phi}^{2} \end{matrix}} \;,
\label{eq:mass_matrices-SU(2)}
\end{align}
with $\lambda_{\Phi}$ given by \eqs{eq:extrema-SU(2)}. It is also evident that, 
as expected, $M_G^2$ becomes a zero matrix (\ie all pNGBs become massless) in the limit of SU(2) invariance.

\subsubsection*{Stability of PNGBs}
In the $U(1)$ case~\cite{Gross:2017dan}, the stability of the DM was a result of a natural dark CP-invariance. Although
it is not possible to absorb all phases of the parameters, here, the PNGBs are still stable. What keeps the PNGBs stable 
is a residual symmetry exhibited by the potential that forbids such mixings between the PNGBs with $\rho$ and $h$.  
To show this, observe that $V_{0}$ is a polynomial of $ |\Phi|^2 = \frac{1}{2} \left(  \chi^{2} +\phi^{2} + s^{2}  +(v_{\Phi}+\rho)^2 \right)$,
which is symmetric under orthogonal rotations of $\lrb{\chi,\phi,s}$, \ie $O(3)$. As a result, no mixing between PNGBs and other scalars can be generated 
from the $SU(2)$ symmetric part of the potential. So, only $V_{\rm soft}$ can induce a mixing between $\lrb{\chi,\phi,s}$ and $\rho$.
However, since $v_{\Phi}$ is always added to $\rho$, any mixing between $\rho$ and a PNGB, induces a linear term (proportional to that PNGB).
That is, all PNGB-$\rho$ mixings should vanish by virtue of the minimization conditions. As an example, by plugging~\eqs{eq:vevs_SU2} in~\eqs{eq:Vsoft_SU(2)},
we get the potential mixing term between $\phi$ and $\rho$
$$
V_{\phi \, \rho}=\dfrac{1}{2}(2 m_{\Phi \, 12}^2 + m_{\Phi \, 21}^{2 \, \prime} + {\rm h.c.})(\rho+v_{\Phi})\phi \;,
$$
which automatically vanishes once we impose~\eqs{eq:extrema-SU(2)}. 
Therefore, by performing an orthogonal rotation to $\lrb{\chi,\phi,s}$ to the 
PNGB eigenvalue basis (with eigenstates $\xi_{1,2,3}$) we see that all $\xi$'s are stable.~\footnote{Also note that there are no  interactions 
that could induce a decay of a PNGB to another (since $V_{0}$ is $O(3)$ symmetric), \eg there are only $\xi_{1}^{2}h$ terms, while interactions 
of the form  $\xi_{1} \xi_{2} h$ are forbidden by the $O(3)$ symmetry of $V_{0}$.}
That is, the Lagrangian is $Z_2^{(\xi_{1})}  \times Z_2^{(\xi_{2})} \times Z_2^{(\xi_{3})}$ symmetric (\ie each PNGB carries its own $Z_{2}$ parity)  
that not only forbids decays of the PNGBs, but also PNGB conversions as well. Thus, all pseudo-Nambu-Goldstone bosons are stable, resulting to a three-component DM 
content.

We note here that, in the limit of decoupled $\phi$ and $s$, we recover the $U(1)$~\cite{Gross:2017dan} case. That is, in this limit, one expects 
(approximately) the same  phenomenology. Thus, the relic abundance of the $SU(2)$ model ($\Omega_{SU(2)} h^{2}$) should be comparable to the  $U(1)$ 
($\Omega_{U(1)}h^{2}$). On the other hand, in the limit of (almost) degenerate PNGBs, the relic abundance should get a factor of 
three, \ie $\Omega_{SU(2)} h^{2} \approx 3\Omega_{U(1)}h^{2}$, which tightens the bound on the
annihilation cross section per DM-particle. This, in turn, means that the required value of the coupling(s) responsible for the DM annihilation should be smaller
(by a factor of $\sim \sqrt{3}$).
In addition to that, the LHC constraints~\cite{Huitu:2018gbc} should remain mostly unaffected, since we would have three degenerate particles each one with 
interactions reduced by a factor of $\sim 3$. Between the degenerate and decoupling limits described above, the picture can get quite involved 
(\eg~\cite{Belanger:2012vp,DiFranzo:2016uzc}). However, in principle
we should expect the relic abundance to be between these two limits, \ie $\Omega_{U(1)}h^{2} \lesssim \Omega_{SU(2)} h^{2} \lesssim 3\Omega_{U(1)}h^{2}$.
Therefore, it seems plausible that there should be some allowed region in the parameter space of the $SU(2)$ case, although a detailed analysis is still needed.

\subsubsection*{The pseudo-Nambu-Goldstone--nucleon interaction}
\begin{figure}
	\centering
	\includegraphics[width=0.7\linewidth]{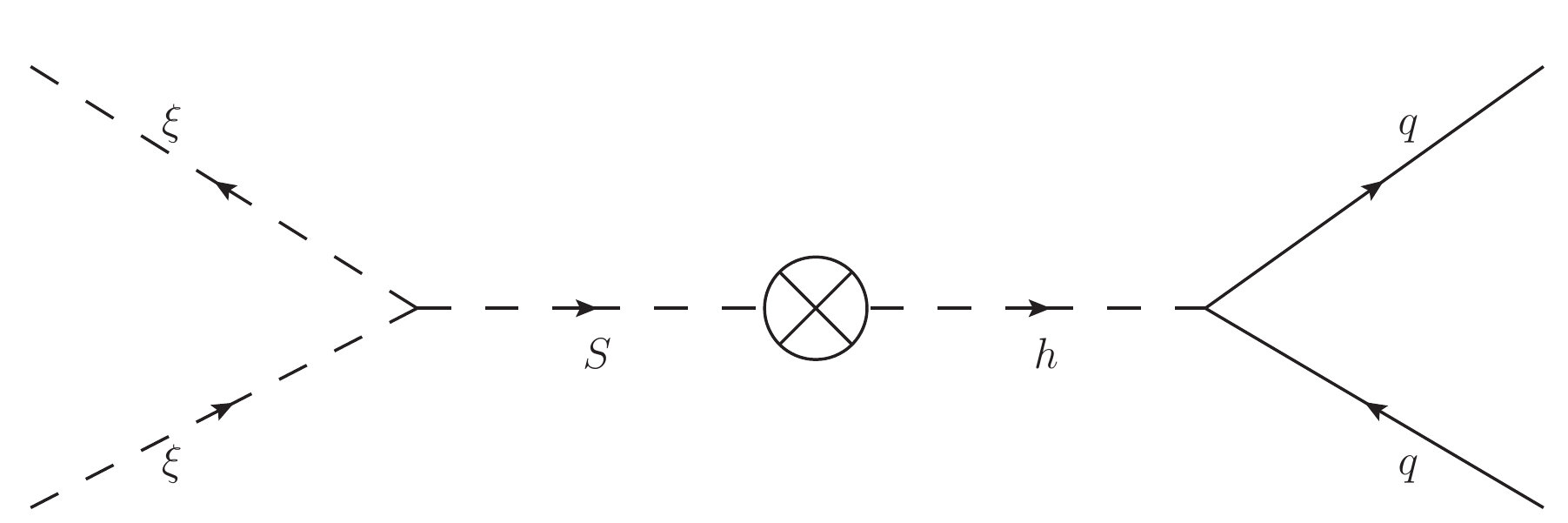}
	\caption{ The Feynman diagram for the elastic scattering  between a quark (q) and a PNGB.}
	\label{fig:Diagram_ADD_N=2}
\end{figure}
Since all PNGBs are stable, we need to calculate three amplitudes for the direct detection cross section. However, due to the $O(3)$ symmetry of the interaction terms, the
amplitude for the $\xi_{i}$-nucleon elastic scattering ($\xi_{i} n \to \xi_{i} n$) is proportional to  $G_{i}$-nucleon elastic scattering amplitude and it is independent of $i$.  
In general, the interaction of the three-point terms pertinent to this interaction can be written as 
\begin{equation}
\mathcal{L}_{\rm int}=-\frac{1}{2}\sum\limits_{i=1}^3\sum\limits_{j=1}^3\sum\limits_{k=1}^2 Y^{(k)}_{ij} G_{i}G_{j}S_{k}\; .
\label{eq:3-point-SU2-general}
\end{equation}
But due to the $O(3)$ symmetry, we expect that
\begin{equation}
\mathcal{L}_{\rm int}=-\frac{1}{2}(G_{1}^{2} + G_{2}^{2} + G_{3}^{2} )\, \sum\limits_{k=1}^2 Y^{(k)} S_{k}=  -\frac{1}{2}(\xi_{1}^{2} + \xi_{2}^{2} + \xi_{3}^{2} )\, \sum\limits_{k=1}^2 Y^{(k)} S_{k}\; .
\label{eq:3-point-SU2-O3}
\end{equation}
From the potential~\ref{eq:V0_SU(2)} and the relations~\ref{eq:extrema-SU(2)},  we obtain
\begin{equation}
\mathcal{L}_{\rm int}=-\frac{1}{2}(\xi_{1}^{2} + \xi_{2}^{2} + \xi_{3}^{2} ) 
\lrb{ \begin{matrix}
	\lambda_{H\Phi} \; v	
	\\
	-  \dfrac{1}{v_{\Phi}}  \lrb{\lambda_{H\Phi} v^2 +4 m_{\Phi_{22}}^{2} +2  m_{\Phi_{22}}^{\prime \, 2}  }
	\end{matrix} }^{T}  
\lrb{ \begin{matrix}
	h	
	\\
	\rho
	\end{matrix} } 
\; .
\label{eq:3-point-SU2-O3}
\end{equation}

Since we are interested in the zero-momentum transfer limit, the propagator is proportional to the inverse of the mass-matrix $M_{S}^{2}$.
Then the direct detection amplitude for all PNGBs (the Feynman diagram is shown in Fig.~\ref{fig:Diagram_ADD_N=2}) becomes   
\begin{equation}
A_{\rm DD} \sim \lrb{ \begin{matrix}
	-\lambda_{H\Phi} v_{\Phi} v
	\\
	\lambda_{H\Phi} v^2 +4 m_{\Phi_{22}}^{2} +2  m_{\Phi_{22}}^{\prime \, 2}  
	\end{matrix} }^{T}
\lrb{\begin{matrix} 
	\lambda_{H\Phi} v^2 +4m_{\Phi_{22}}^{2}+2m_{\Phi_{22}}^{\prime \, 2}  & \lambda_{H\Phi}  v_{\Phi}v  \\
	\lambda_{H\Phi}  v_{\Phi}v & -\lambda_{H} v^2 \end{matrix} }
\lrb{ \begin{matrix}	1 	\\ 	0	\end{matrix} }=0
\;,
\label{eq:A_DD-SU(2)}
\end{equation}
which concludes the proof of the claim that the DM-nucleon cross section vanishes at tree-level and zero momentum transfer. 
However, this only indicates that the direct detection cross section is ``naturally suppressed". In practice,  loop corrections need to be included  as well, since these effects
could allow for a possible direct detection signal~\cite{Gross:2017dan,Azevedo:2018exj,Ishiwata:2018sdi}.

\subsubsection*{Generalization to $SU(N)$}

It is straightforward to generalize the above result in the case where $\Phi$ is in the fundamental 
representation of a softly broken $SU(N)$ global symmetry, since the form of  $V_0$ is the same 
as in \eqs{eq:V0_SU(2)}, with the soft breaking terms being
\begin{equation}
V_{\rm soft}= \sum\limits_{i=1}^N \sum\limits_{j=1}^N \Big[ m_{\Phi \, ij}^2 \Phi_{i} \Phi_{j}+{\rm h.c.} + m_{\Phi \, ij}^{\prime \, 2} \Phi^{\dagger}_{i} \Phi_{j} \Big] \;,
\label{eq:Vsoft_SU(N)}
\end{equation}
where in  analogy to $SU(2)$ $m_{\Phi \, ij}=m_{\Phi \, ji}$ and $m_{\Phi \, ij}^{\prime}= m_{\Phi \, ji}^{\prime \, *}$.
Assuming that the $N^{\rm th}$ component of $\Phi$ develops a VEV, one can show  that the minimization of the potential requires
\begin{align}
&m_{\Phi \, NN}^2  \in  \mathbb{R} \nonumber \\
&\mu_{H}^2 =  \lambda_{H} v^2 + \lambda_{H\Phi} v_{\Phi}^2 \nonumber \\ 
&m_{\Phi \, iN}^{2} = - \frac{m_{\Phi \, Ni}^{\prime \, 2}}{2}, \quad \forall i<N \nonumber \\
&\lambda_{\Phi} =  -\frac{1}{v_{\Phi}^2}\left[\lambda_{H \Phi}v^2 + 4 m_{\Phi_{NN}}^2 + 2 m_{\Phi_{NN}}^{\prime \, 2}  \right] \;,
\label{eq:extrema-SU(N)}
\end{align} 
This results to $2N-1$ PNGBs, $\chi$, $\phi_{i}$  and $s_{i}$ with $i=1,2, \dots, N-1$,
where, in complete analogy to the $SU(2)$ case, the interaction potential ($V_0$) becomes symmetric under
$O(2N-1)$,\footnote{The symmetric potential depends on $|\Phi|^{2} \sim 
\lrb{ \chi^{2} +\phi_{1}^{2} + s_{1}^{2} + \phi_{2}^{2} + s_{2}^{2}+ \dots \phi_{N-1}^{2} + s_{N-1}^{2} }+(v_{\Phi}+\rho)^2$, which is symmetric under orthogonal rotations of the PNGBs.} which results to a $Z_2^{(\xi_{1})}  \times Z_2^{(\xi_{2})}  \dots \times Z_2^{(\xi_{2N-1})}$ symmetry for the 
entire potential. Therefore, all  pseudo-Nambu-Goldstone bosons are stable particles. We also point out that the same arguments for the
relic abundance of the $SU(2)$ case hold also in $SU(N)$. That is, in general the relic abundance should be 
$\Omega_{U(1)} h^{2} \lesssim \Omega_{SU(N)}h^{2} \lesssim  (2N-1) \Omega_{U(1)} h^{2}$.
Returning to the discussion for the direct detection cross section,  the pseudo-Nambu-Goldstone boson--nucleon interaction terms take the familiar form

\begin{equation}
\mathcal{L}_{\rm int}=-\frac{1}{2}\sum\limits_{i=1}^{2N-1} \xi_{i}^{2}  \lrb{ \begin{matrix}
	\lambda_{H\Phi} \; v	
	\\
	-  \dfrac{1}{v_{\Phi}}  \lrb{\lambda_{H\Phi} v^2 +4 m_{\Phi_{22}}^{2} +2  m_{\Phi_{22}}^{\prime \, 2}  }
	\end{matrix} }^{T}  
\lrb{ \begin{matrix}
	h	
	\\
	\rho
	\end{matrix} }  \; ,
\label{3-point-general}
\end{equation}
Since the mass matrix $M_{S}^2$ is independent of $N$ (\ie it is always given by~\eqs{eq:Mass_matrices-SU(2)}),
the  amplitude for the process $\xi_{i} N \to \xi_{i} N$ at tree-level and zero momentum transfer, vanishes as in the $SU(2)$ case. 

One should keep in mind that the cancellation takes place only if $\Phi$ is in the fundamental representation of $SU(N)$.
It is not clear if  $A_{\rm DD}$ would cancel if another (irreducible) representation of $\Phi$ was assumed, as there are additional
interactions, corresponding to all the possible contractions of the $SU(N)$ indices.
For example, for $N=2$ and $\Phi$ in the adjoint representation, there is an interaction term of the form

\begin{equation*}
V_{\rm int} \sim |H|^2 \sum\limits_{i,j,k,l} \epsilon^{il}\epsilon^{jk} \Phi_{ij}\Phi_{kl},
\end{equation*}
which can potentially change the mixing between the particles in a non-trivial way. Since
the number of such interactions increases greatly with the dimension of each representation
of $SU(N)$, it becomes hard to generalize. Thus, we postpone such analysis for the future.~\footnote{
However, if the imposed symmetry is $SU(N) \times U(1)$, such interactions are not allowed,
which means the mechanism under consideration holds for other representations as well.}

\subsubsection*{Beyond the tree-level approximation}
\begin{figure}[h!]
	\centering
	\includegraphics[width=0.85\linewidth]{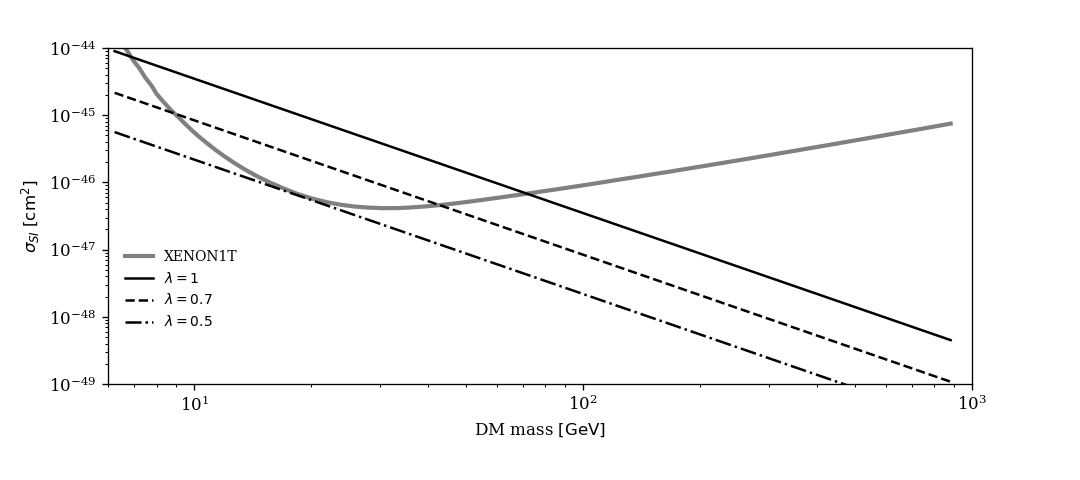}
	\caption{An estimate of the 1-loop direct detection cross section~eq.~(\ref{eq:sigma_SI})  for $\lambda=1$ (black line), $\lambda=0.7$ (dashed line), $\lambda=0.5$ (dashed-dotted line). The gray line corresponds to the 
		upper limit as given by XENON1T~\cite{Aprile:2018dbl}.  }
	\label{fig:sigma_SI}
\end{figure}
So far, we have considered the direct detection cross section at the tree-level, which vanishes because of 
the PNGB nature of the DM particles, \ie the approximate imposed symmetry. However, one expects that 
new interaction terms can be induced at the loop-level, from the contribution of the soft breaking terms.
That is, one expects four-point symmetry breaking interaction terms (\eg $ |H|^2 \Phi_{1}\Phi_{1} +\mathrm{h.c.}$) to be generated.~\footnote{Note that three-point interactions
cannot be produced because the entire potential is symmetric under $\Phi_{1,2} \to -\Phi_{1,2}$.}  
In this case, we expect a situation similar to ref.~\cite{Gross:2017dan}, with a typical loop-induced coupling
$\lambda^{\prime} \sim \frac{\lambda^{2}}{(4\pi)^{2}}$ (multiplied by a function logarithmic in the mass parameters as in~\cite{Gross:2017dan}, which should vanish in the symmetric limit), 
and $\lambda^{2}$ proportional to  a combination of $\lambda_{H \Phi} \lambda_{\Phi}+\lambda_{\Phi}^{2}$. 
That is, one expects for the direct detection cross section to be suppressed. 
An order of magnitude estimate can be deduced by assuming an interaction of the form (the tree level coupling cancels at zero momentum transfer, 
so we only show the loop-induced one here) $\mathcal{L}_{h \xi \xi } \sim \frac{\lambda^{2}}{(4\pi)^{2}}\, v \, h  \, \xi^{2}$ 
($\xi$ is a DM particle).~\footnote{Note that such interaction, in our case, are also multiplied by the mixing of $\rho$ and $h$. By omitting them, we may overestimate the 
DM-nucleon cross section.} For such interactions, the spin-independent cross section is approximately~\cite{Hardy:2018bph}
\begin{equation}
\sigma_{SI} \sim  3.5 \times 10^{-47} \left(\dfrac{100 \, \GeV}{m_{\xi}}\right)^{2}  \lambda^{4} \, {\rm cm}^{2}   \, ,
\label{eq:sigma_SI}
\end{equation}
which for a moderate $\lambda \sim 0.5$ is below current limits (see fig.~\ref{fig:sigma_SI}). 
Even for $\lambda =1 $, this cross section is below the bounds, for most of the DM mass range.
Note that since ~\eqs{eq:extrema-SU(N)} suggests that 
$\lambda^{2} \sim \lambda_{H \Phi} \lambda_{\Phi}+\lambda_{\Phi}^{2} \sim \frac{\lambda_{H \Phi}^{2} v^2}{v_{\Phi}^{2}} \lrsb{ 1-\lrb{\frac{v}{v_{\Phi}}}^{2} + \dots }$, such values of $\lambda$ are 
reasonable, assuming $v_{\Phi}>v,m_{\Phi \,NN}^{(\prime)}$.
In principle, though, since we may have omitted important loop factors, one  should calculate the relevant one-loop vertices ( or the complete 1-loop scalar potential as also stated in~\cite{Gross:2017dan}),  in order to have an accurate description of these interactions. 
Finally, one should keep in mind, that in the case of multi-component DM, each component contributes to the direct detection cross section according to 
its relative relic abundance~\cite{DiFranzo:2016uzc}. This could mean more relaxed direct detection bounds (if the DM masses are separated), since for the 
various DM components, the DM-nucleon  cross section should be rescaled as $\sigma^{i}_{SI} \approx \sigma_{SI} \times \frac{\Omega^{i}h^{2}}{\Omega^{tot}h^{2}}$ (for the $i^{\rm th}$ component).

\subsubsection*{Note on possible completions}
The models presented here should not be considered UV-complete, since the origin of the
global symmetries as well as the soft breaking terms are not known. That is, such models should be 
treated as low-energy limits of other, UV-complete, ones.
Possible UV-completions, may include new gauge symmetries and a complicated spectrum of particles, so the
explicitly broken symmetries may be manifested as approximate symmetries (as the so-called ``custodial symmetry" in
the SM~\cite{Sikivie:1980hm}) in their low-energy limit.
However, the structure of the low-energy models should be mostly unaffected, and any new
effects induced by the completion, should be suppressed by some characteristic high-energy scale.
In principle, such completion can induce decays of the DM particles, as well as other effects 
(\eg tree-level DM interaction with the nucleons) not present in the low-energy model,
but suppressed by the energy scale of the completion.~\footnote{For example, in ref.~\cite{Gross:2017dan} the DM particle
becomes unstable with a lifetime of order $10^{39} \; {\rm s}$.} Such effects, though, can only be studied in a case-by-case manner 
provided a valid completion.

\section{The $U(1)\times S_2$ case}\label{sec:S_N}
\setcounter{equation}{0}
In this section, we examine another case, which we denote as $U(1)\times S_2$. In this case, the SM is extended 
by two scalars ($S_{1,2}$) charged only under a softly broken global $U(1)$.  For the desired cancellation to occur, we impose a permutation symmetry on  $S_{1,2}$. 
As we will see, this symmetry  provides a sufficient condition for the vanishing of the PNGB-nucleon cross section. 

\subsection{The cancellation mechanism for this model}\label{sec:U(1)*S_2}
\subsubsection*{The Potential}
In the case of two scalars, each transforming as $S_{i} \to e^{-i a}S_{i}$, the  $U(1) \times S_{2}$  symmetric potential,
assuming that all parameters are real numbers (we shall call this assumption dark CP--invariance), is
\begin{align}
	V_{0}&=-\frac{\mu_{H}^2}{2}    \left| H \right|^{2}  \, + \, \frac{\lambda_{H}^2}{2}   \left| H \right|^{4}  \, + \, 
	\lambda_{HS_{1}}   \left| H \right|^{2}  \lrb{ \left| S_{1} \right|^{2} +\left| S_{2} \right|^{2}   }  \, + \,
	\lambda_{HS_{2}}   \left| H \right|^{2} \lrb{ S_{1}S_{2}^{\dagger}	+ {\rm h.c.} }   \nonumber  \\  
	&\, - \, \frac{\mu_{S_{1}}^2}{2}    \lrb{ \left| S_{1} \right|^{2} +\left| S_{2} \right|^{2}   }
	 - \,  \frac{\mu_{S_{2}}^2}{2} \lrb{ S_{1}S_{2}^{\dagger}	+ {\rm h.c.} } \, + \,
	\frac{\lambda_{S_{1}}}{2}   \lrb{   \left| S_{1} \right|^{4} +  \left| S_{2} \right|^{4}  } \, + \,
	\frac{\lambda_{S_{2}}}{2}  \lrsb{ (S_{1}S_{2}^{\dagger} )^{2}	+ {\rm h.c.} } \nonumber \\
	&+ \,  \lambda_{S}^{\prime} |S_{1}|^2|S_{2}|^2  
	+ \, c \,  \lrb{  S_{1} \, S_{2}^{\dagger} + S_{2} \, S_{1}^{\dagger}} \lrb{ \left|S_{1}\right|^2 + \left|S_{2}\right|^2} 
	  \; ,
	\label{eq:V0_N=2}
\end{align}
while the $S_{2}$-symmetric soft breaking potential  is written as
\begin{equation}
V_{\rm soft}=-  \frac{\mu_{S_{1}}^{\prime \, 2}}{2} \lrb{ S_{1}^{2} + S_{2}^{2} + \mathrm{h.c.} } \, - \,  \mu_{S_{2}}^{\prime \, 2} \lrb{ S_{1} S_{2} + \mathrm{h.c.} }  \;.
\label{eq:Vsoft_N=2}
\end{equation}
with the total potential given by  $V=V_{0}+V_{\rm soft}$. In order to find the minimization conditions, we expand the fields around  their VEVs
\begin{eqnarray}
& S_{1,2} = \frac{1}{\sqrt{2}}(v_S +s_{1,2} +i \, \chi_{1,2}) \; , \nonumber\\
& H = \frac{1}{\sqrt{2}}\left(\begin{matrix}0 \\ v+h \end{matrix}\right)\; ,
\label{eq:vevs_N=2}
\end{eqnarray}
where this particular choice of  $\vev{S_{1,2} }$, ensures that the potential remains symmetric under simultaneous permutations of $\left( s \, , \,  \chi \right)_{1} \leftrightarrow \left( s \, , \,  \chi \right)_{2}$.
Due to the permutation symmetry, there are only two independent stationary point conditions, which read
\begin{eqnarray}
\mu_{H}^{2}  &=& \lambda_{H}  v^{2} +2 v_S^{2} \lrb{\lambda_{HS_{1}}+ \lambda_{HS_{2} }} \, , \\
\mu_{S_{1}}^2 &=&v^2\lrb{\lambda_{HS_{1}}+ \lambda_{HS_{2}}  } \, - \, \lrsb{ \mu_{S_{2}}^2+ 2 \lrb{ \mu_{S_{1}}^{\prime \; 2} +  \mu_{S_{2}}^{\prime \; 2} } }   \nonumber \\
& &+v_{S}^2 \lrb{
	\lambda_{S_{1}}+  \lambda_{S_{2}}+ \lambda_{S}^{\prime} +4 c }\, .\nonumber 
\label{eq:minimization_N=2}
\end{eqnarray}

\subsubsection*{Spectrum of the CP--odd scalars}
In order to calculate the direct detection amplitude, we first need to identify the PNGB. This can be done by diagonalizing the mass matrix of the CP--odd fields to its eigenvalues.
Once the eigenvalues are found, one of them should vanish in the limit where the $U(1)$ is restored, which should correspond to the PNGB. From \eqs{eq:minimization_N=2}, we 
obtain the mass matrix for the $\chi$'s

\begin{equation}
M_{\chi}^{2}=\lrb{
\begin{matrix}
-   \frac{v^2\lambda_{HS_{2}}}{2}- v_{S}^2 (\lambda_{S_{2}} + c ) +2 \mu_{S_{1}}^{\prime\; 2} +  \mu_{S_{2}}^{\prime \; 2} + \frac{\mu_{S_{2}}^2}{2} &
   \frac{v^2\lambda_{HS_{2}}}{2} + v_{S}^2 (\lambda_{S_{2}} + c ) + \mu_{S_{2}}^{\prime\; 2}  - \frac{\mu_{S_{2}}^2}{2} \\
   \frac{v^2\lambda_{HS_{2}}}{2} + v_{S}^2 (\lambda_{S_{2}} + c ) + \mu_{S_{2}}^{\prime\; 2}  - \frac{\mu_{S_{2}}^2}{2} &  
-   \frac{v^2\lambda_{HS_{2}}}{2}- v_{S}^2 (\lambda_{S_{2}} + c ) +2 \mu_{S_{1}}^{\prime\; 2} +  \mu_{S_{2}}^{\prime \; 2} + \frac{\mu_{S_{2}}^2}{2}
\end{matrix}},
\label{eq:Mx_N=2}
\end{equation}
from which we find the eigenvalues
\begin{eqnarray}
m_{\xi_{1}}^{2}&=& 2(\mu_{S_{1}}^{\prime \, 2} + \mu_{S_{2}}^{\prime \, 2}) \nonumber \\
m_{\xi_{2}}^{2}&=& 2  \mu_{S_{1}}^{\prime \; 2} + \mu_{S_{2}}^2-v^2 \lambda_{HS_{2}} - 2 v_{S}^2 (\lambda_{S_{2}} + c ) \; .
\label{eq:Mxi_N=2}
\end{eqnarray}
It is apparent that $m_{\xi_{1}}^{2}$ vanishes in the limit $\mu_{S_{1,2}}^{\prime \, 2} \to 0$, thus the particle corresponding to this mass can be identified as the 
would-be Nambu-Goldstone boson of the $U(1)$, \ie the PNGB of this model. The eigenstates corresponding to these masses are 
\begin{eqnarray}
\xi_{1}=\dfrac{1}{\sqrt{2}}(\chi_{1}+\chi_{2}) \, , \,
\xi_{2}=\dfrac{1}{\sqrt{2}}(\chi_{1}-\chi_{2})\; .
\end{eqnarray}
It is worth noting that the PNGB ($\xi_{1}$) is symmetric under $\chi_{1} \leftrightarrow \chi_{2}$. This property of the PNGB, will be proven helpful especially in the $N$-particle
generalization  of this model, since it will allow us to calculate the desired direct detection amplitude easily.  
The imposed dark CP--invariance can potentially keep both of the states $\xi_{1,2}$ stable, since there are only interactions involving even numbers of CP--odd particles, \eg there is 
no $\xi_{1} \, h^{2}$ interaction term while the vertex $\xi_{1}^{2} \, h$ exists. However, since we are interested in the scenario where the DM particle is a PNGB, we need to impose an extra
hierarchy condition, so that $\xi_{1}$ will be stable while $\xi_{2}$ will be able to decay.   
This condition is $m_{\xi_{1}}<m_{\xi_{2}}$, with their difference ($m_{\xi_{2}}-m_{\xi_{1}}$) at least larger than the mass of the lightest CP--even particle (\eg $m_{\xi_{2}}-m_{\xi_{1}}>m_{H} \approx 125 \GeV $ if the Higgs boson is the lightest one). 
This is not too restrictive, and it does not affect the vanishing of the PNGB-nucleon cross section, but it must be pointed out for the sake of completeness. Also, as in sec.~\ref{sec:SU(2)}, it seems 
reasonable that this model will be allowed by observations, at least close to the limit in which becomes similar to the $U(1)$ case. That said, however, since the parameter space is greater here, 
there should be room to accommodate all constraints (especially since the direct detection bounds are evaded).
 

\subsubsection*{The direct detection amplitude}
The calculation of the quark-$\xi_{1}$ scattering amplitude is a relatively straightforward task. We just need to calculate the corresponding Feynman diagram 
(fig.~\ref{fig:Diagram_ADD_N=2}). In fact, since we are interested in the zero momentum transfer limit,  the  ingredients that we need in order to show that the direct detection cross section vanishes, are the inverse of the mass matrix of the CP--even scalars and the three-point interaction of a pair of PNGBs with them (\ie vertices of the 
form $\xi_{1}^2 \, h$ and $\xi_{1}^2 \, s_{1,2}$). 
The mass terms for the CP--even scalars can be written in a compact form as
\begin{equation}
\mathcal{L}_{\rm hs}=-\frac{1}{2} \Phi^{T} \, M^2_{\Phi} \, \Phi \; , 
\end{equation}
with  $\Phi= \lrb{ h, s_{1}, s_{2} }^{T}$ and 
\begin{equation}
M_{\Phi}^{2}=\lrb{
\begin{matrix}
v^2 \lambda_{H} & 
v v_{S} (\lambda_{HS_{1}} + \lambda_{HS_{2}}) & 
v v_{S} (\lambda_{HS_{1}} + \lambda_{HS_{2}})  \\[0.1cm]
v v_{S} (\lambda_{HS_{1}} + \lambda_{HS_{2}})  & 
\frac{ 2v_{S}^2 (c+\lambda_{S_{1}} )+\mu_{S_{2}}^{2}+2\mu_{S_{2}}^{\prime \, 2}-\lambda_{HS_{2}}v^{2}}{2}    & 
 \frac{2v_{S}^{2} (3 c + \lambda_{S_{2}} + \lambda_{S_{1}}^{\prime} )+v^{2} \lambda_{ HS_{2}} - \mu_{S_{2}}^{2} - 2 \mu_{S_{2}}^{\prime \, 2}}{2} \\[0.1cm]
v v_{S} (\lambda_{HS_{1}} + \lambda_{HS_{2}})  & 
 \frac{2v_{S}^{2} (3 c + \lambda_{S_{2}} + \lambda_{S_{1}}^{\prime} )+v^{2} \lambda_{ HS_{2}} - \mu_{S_{2}}^{2} - 2 \mu_{S_{2}}^{\prime \, 2}}{2}  &
\frac{ 2v_{S}^2 (c+\lambda_{S_{1}} )+\mu_{S_{2}}^{2}+2\mu_{S_{2}}^{\prime \, 2}-\lambda_{HS_{2}}v^{2}}{2}
\end{matrix}
}
\end{equation}
Observing that only $h$ couples to SM fermions, we only need the following few terms of the inverse of $M_{\Phi}^{2}$

\begin{align}
\lrsb{M_{\Phi}^2}^{-1}_{11}& \sim v_{S}^{2}\lrb{\lambda_{HS_{1}}+ \lambda_{HS_{2}}+ \lambda_{HS_{1}}^{\prime}+4c}  \label{eq:Prop_N=2}\\
\lrsb{M_{\Phi}^2}^{-1}_{i1} & \sim \, -v v_{S} (\lambda_{HS_{1}} + \lambda_{HS_{2}})  \, . \nonumber
\end{align} 

With the interaction term of the Lagrangian terms responsible for the $\xi_{1}$-nucleon elastic scattering being ~\footnote{Note that since $\xi_{1}$
is an $S_{2}$ symmetric state, a pair of $\xi_{1}$ interacts in the same way with both  $s_{1}$ and $s_{2}$.} 
\begin{equation}
\mathcal{L}_{\rm int} =- \frac{1}{8} \xi_{1}^2 \lrb{ \begin{matrix}
	 2 v (\lambda_{HS_{1}} + \lambda_{HS_{2}}) \\
	v_{S} \lrb{\lambda_{HS_{1}}+ \lambda_{HS_{2}}+ \lambda_{HS_{1}}^{\prime}+4c} \\
	v_{S} \lrb{\lambda_{HS_{1}}+ \lambda_{HS_{2}}+ \lambda_{HS_{1}}^{\prime}+4c}
\end{matrix} }^{T}
 \lrb{\begin{matrix} h \\ s_{1} \\ s_{2} \end{matrix} }, \;\;
\label{eq:L_int_N=2}
\end{equation}
we can show that the  the amplitude for the $\xi_{1}$-nucleon elastic scattering vanishes. That is

\begin{equation}
A_{\rm DD}\sim \lrb{ \begin{matrix}
	2 v (\lambda_{HS_{1}} + \lambda_{HS_{2}}) \\
	v_{S} \lrb{\lambda_{HS_{1}}+ \lambda_{HS_{2}}+ \lambda_{HS_{1}}^{\prime}+4c} \\
	v_{S} \lrb{\lambda_{HS_{1}}+ \lambda_{HS_{2}}+ \lambda_{HS_{1}}^{\prime}+4c}
	\end{matrix} }^{T} 
\lrb{ \begin{matrix}
	v_{S}^{2}\lrb{\lambda_{HS_{1}}+ \lambda_{HS_{2}}+ \lambda_{HS_{1}}^{\prime}+4c} \\
	-v v_{S} (\lambda_{HS_{1}} + \lambda_{HS_{2}})  \\
	-v v_{S} (\lambda_{HS_{1}} + \lambda_{HS_{2}}) 
	\end{matrix} }=0\;. \label{eq:ADD_N=2}
\end{equation}  
%

In analogy to the $SU(2)$ case, one again expects one-loop correction. This correction should be suppressed, with an induced coupling 
$\sim \frac{\lambda^2}{(4\pi)^2}$, with $\lambda$ combination of all couplings in this model. The case here is more involved, however, 
since the number of independent  parameters is greater.

\subsection{Generalization to $U(1)\times S_N$}\label{sec:U(1)*S_N}
As we saw in sec.~\ref{sec:U(1)*S_2}, the cancellation mechanism holds when the model consists of two scalars under the assumption that the potential is symmetric 
under permutations of these scalars. This  symmetry fixes the PNGB-$s_{1,2}$ interactions  and the relevant components of  $M_{\Phi}^{2}$ in such way 
that $A_{DD}$ vanishes. However, there is no guarantee that this also happens if we add more scalars, since more interaction terms 
are allowed. In this section, we investigate whether $A_{DD}$ vanishes in a model consisting of an arbitrary number of scalars. We denote this model as 
$U(1) \times S_{N}$, and it is a direct generalization of $U(1) \times S_{2}$ with $N$ number of scalars. 

\subsubsection*{The Potential for $N$ Scalars}
In the case of $N$ scalar fields, each transforming as $S_{i} \to e^{-i a}S_{i}$ (similarly to sec.~\ref{sec:U(1)*S_2}), the  $U(1)\times S_N$  
symmetric potential, assuming again dark CP--invariance, can be written as

\begin{eqnarray}
V_{0}&=-\frac{\mu_{H}^2}{2} |H|^{2} \, + \, \frac{\lambda_{H}^2}{2} |H|^{4} - 
\sum\limits_{i,j}  \frac{\mu_{S_{ij}}^2}{2}  S_{i} S_{j}^{\dagger} +
\sum\limits_{i,j}\frac{\lambda_{S_{ij}}}{2} (S_{i}S_{j}^{\dagger})^2 +
 \sum\limits_{i,j} \frac{\lambda_{S_{ij}}^{\prime}}{2} |S_{i}|^2|S_{j}|^2 \nonumber \\
&+ \sum\limits_{i,j,k} c_{ijk} S_{i}S_{j}^{\dagger} |S_{k}|^2+\sum\limits_{i,j,k} c^{\prime}_{ijk} (S_{i}S_{j} S_{k}^{\dagger \, 2 }  + {\rm h.c.})
\, + \, \sum\limits_{i,j,k,l}d_{ijkl} S_{i}S_{j} S_{k}^{\dagger}S_{l}^{\dagger} \nonumber \\
&+\sum\limits_{i,j}\lambda_{HS_{ij}} |H|^2S_{i}S_{j}^{\dagger}\; , 
\label{eq:V0_N}
\end{eqnarray}
where all the sums run over all scalars. This potential has some redundant terms, so we can set some of them to zero:
\begin{align}
\lambda_{Sii}^{\prime}&=0  \nonumber\\
c_{iik}&=0  \nonumber\\
 c_{iij}^{\prime}&=c_{iji}^{\prime}=c_{ijj}^{\prime}=0 \nonumber \\
d_{iijk}&=d_{ijik}=d_{ijki}=d_{ijjk}=d_{ijkj}=d_{ijkk}=0 \; . \label{eq:zeros_SN}
\end{align} 
Furthermore, the permutation symmetry, dictates:
\begin{eqnarray}
&\mu^2_{S_{ij}}=&		 	\hspace{-0.2cm} \Big\{ \begin{matrix} \mu^2_{S_{1}} &  ( i=j )\\ \mu^2_{S_{2}} & ( i \neq j) \end{matrix}       \nonumber   \\
&\lambda_{S_{ij}}=&	\hspace{-0.2cm}	\Big\{ \begin{matrix} \lambda_{S_{1}} & ( i=j ) \\ \lambda_{S_{2}} & ( i \neq j )\end{matrix}            \nonumber      \\
&\lambda_{HS_{ij}}=	&	\hspace{-0.2cm}  \Big\{ \begin{matrix} \lambda_{HS_{1}} &  \hspace{-0.25cm}( i=j ) \\ \lambda_{HS_{2}} &   \hspace{-0.25cm} ( i \neq j )\end{matrix}   \nonumber  \\
&\lambda_{S_{ij}}^{\prime}=&	  \hspace{-0.2cm} \lambda_{S}^{\prime} \hspace{0.7cm}  ( i \neq j ) \label{eq:conditions_SN} \\
&c_{iji}=c_{jii}=&			\hspace{-0.2cm}		c_{1}  \; \hspace{0.7cm} (i \neq j  )  \nonumber   \\
&c_{ijk}=	&	 		\hspace{-0.2cm}			c_{2}  \hspace{0.8cm} ( i \neq j \neq k )   \nonumber   \\
&c_{ijk}^{\prime}=&			\hspace{-0.2cm}		c^{\prime}  \hspace{0.85cm} (\  i \neq j \neq k   )  \nonumber  \\
&d_{ijkl}=&								\hspace{-0.2cm}		d  	\hspace{0.9cm}	( i \neq j \neq k \neq l ) \; .  \nonumber  
\end{eqnarray}

As previously, we assume soft breaking of $U(1)$. That is, we add the following terms in the potential
\begin{equation}
V_{\rm soft}=- \sum\limits_{i,j} \frac{\mu_{Sij}^{\prime \, 2}}{2} S_{i}S_{j}+ \mathrm{h.c.}\;,
\label{Vsoft_N}
\end{equation}
where, due to the $S_N$ symmetry, we have 
\begin{equation}
\mu^{\prime \; 2}_{S_{ij}}=		 \Big\{ \begin{matrix} \mu^{\prime \; 2}_{S_{1}} &  ( i=j )\\[0.15cm] \mu^{\prime \; 2}_{S_{2}} & ( i \neq j) \end{matrix}  
\label{eq:soft_breaking-S_N-relations}
\end{equation}
So, from \eqs{eq:zeros_SN,eq:conditions_SN,eq:soft_breaking-S_N-relations}, the total potential becomes
\begin{eqnarray}
V &=-\dfrac{\mu_{H}^2}{2} |H|^{2} \, + \, \dfrac{\lambda_{H}^2}{2} |H|^{4} +\lambda_{HS_{1}} \sum\limits_{i} |H|^2 |S_{i}|^{2} +\lambda_{HS_{2}} \sum\limits_{i \neq j} |H|^2 S_{i} S_{j}^{\dagger} 
\nonumber \\
&-\dfrac{\mu_{S_{1}}^2}{2} \sum\limits_{i}    |S_{i}|^{2} + \dfrac{\mu_{S_{2}}^2}{2} \sum\limits_{i \neq j}    S_{i} S_{j}^{\dagger} 
+\dfrac{\lambda_{S_{1}}}{2} \sum\limits_{i } |S_{i}|^{4} + \dfrac{\lambda_{S_{2} }}{2}  \sum\limits_{i \neq j}(S_{i}S_{j}^{\dagger})^2 
\nonumber  \\ 
&+ c_{1} \sum\limits_{i \neq j} \lrb{ S_{i} S_{j}^{\dagger} |S_{j}|^{2} + S_{i} S_{j}^{\dagger} |S_{i}|^{2}  } + 
c_{2}   \hspace{-0.2cm} \sum\limits_{  i \neq j \neq k	 }  \hspace{-0.2cm} S_{i} S_{j}^{\dagger} |S_{k}|^{2}  
\nonumber  \\
& + \lambda_{S}^{\prime} \sum\limits_{j > i} |S_{i}|^2 |S_{j}|^2 +  c^{\prime} \hspace{-0.2cm} \sum\limits_{  i \neq j \neq k  }  \hspace{-0.2cm} (S_{i}S_{j} S_{k}^{\dagger \, 2 }  + {\rm h.c.}) 
 + d \hspace{-0.3cm} \sum\limits_{ i \neq j \neq k \neq l} \hspace{-0.3cm} S_{i}S_{j} S_{k}^{\dagger}S_{l}^{\dagger}
\nonumber \\
&-\frac{\mu_{S_{1}}^{\prime \, 2}}{2} \sum\limits_{i}  \lrb{S_{i}^2+ \mathrm{h.c.}} -\mu_{S_{2}}^{\prime \, 2} \sum\limits_{i} \lrb{ S_{i} S_{j} + \mathrm{h.c.}} \;. 
\label{eq:V_N}
\end{eqnarray}
At this point, it becomes clear that the $S_N$ symmetry helps keeping the number of new free parameters relatively small.~\footnote{ There are $10, \; 12 $, and $13$ free parameters 
for $N=2, \; N=3 $, and $N \geq 4$, respectively.} This keeps the model as simple as possible, considering the potential large number of particles.

Similar to the previous, the scalars acquire VEVs
\begin{eqnarray}
& S_{i} = \frac{1}{\sqrt{2}}(v_S +s_{i} +i \, \chi_{i}) \; , \nonumber\\
& H = \frac{1}{\sqrt{2}}\left(\begin{matrix}0 \\ v+h \end{matrix}\right)\; ,
\label{eq:vevs_N}
\end{eqnarray}
where, again, we have assumed that the potential remains symmetric under  $\left( \begin{matrix} s ,\; \chi \end{matrix} \right)_{i} \leftrightarrow 
\left( \begin{matrix} s ,\; \chi \end{matrix} \right)_{j}$ after SSB.
From \eqs{eq:V_N,eq:vevs_N}, we observe that there are only two independent stationary point conditions, due to the $S_{N}$ symmetry, (similar to \ref{sec:U(1)*S_2}),
which  are  
\begin{eqnarray}
\mu_H^2&=&\lambda_H v^2 +N v_{S}^{2}  \lrsb{ \lambda_{HS_{1}}+(N-1) \lambda_{HS_{2}}  } \; , \\
\mu_{S_{1}}^2&=&v^2 \lrsb{\lambda_{HS_{1}}+(N-1) \lambda_{HS_{2}}} - \lrsb{2 \mu_{S_{1}}^{\prime \; 2} + (N-1)(\mu_{S_{2}}^2+2\mu_{S2}^{\prime \; 2}) } \nonumber \\
&+&v_{S}^2 \lrBigsb{
\lambda_{S_{1}}+(N-1)( \lambda_{S_{2}}+ \lambda_{S}^{\prime} +4 c_{1})+
2(N-1)(N-2)(c_{2}+2c^{\prime})\nonumber\\
&+&2(N-1)(N-2)(N-3)d
}\;.\nonumber 
\label{minimization_N}
\end{eqnarray}
These conditions further reduce the number of new parameters by one, \ie the maximum number of new parameters introduced is $12$ for $N \geq 4$ (for $N=2$ and $3$ these are $9$ and $11$, respectively). 

\subsubsection*{Spectrum of the CP--odd scalars}
As in sec.~\ref{sec:U(1)*S_2}, our next step is to find which mass eigenstate corresponds to the PNGB. To do so, we first have to find the mass matrix ($M_{\chi}^{2}$) for the CP--odd scalars. 
Since the CP--odd and CP--even scalars do not mix (due to the dark CP--invariance), their mass terms are symmetric under permutations of the $\chi$'s.
As a result, there are only two different entries in the mass matrix for $\chi$'s, the diagonal, $\lrb{M_{\chi}^2}_{ii}$, and the off diagonal, $\lrb{M_{\chi}^2}_{ij}$, ones.
After some algebra, one can show that
\begin{align}
[M_{\chi}^2]_{ii}&= -  v^2 (N - 1)\frac{\lambda_{HS_{2}}}{2} +2 \mu_{S_{1}}^{\prime\; 2}+ \frac{1}{2}(N - 1)(2\mu_{S_{2}}^{\prime \; 2} + \mu_{S_{2}}^2) \nonumber \\
&- v_S^2 \lrBigsb{
 (N - 1) (\lambda_{S_{2}} + c_{1}) + 
 \frac{1}{2}(N - 1) (N - 2) (c_{2}+ 6 c^{\prime}) \\
&+ (N - 1) (N - 2) (N - 3) 
}  \;,\nonumber \\
[M_{\chi}^2]_{ij}&=\frac{2 \lrb{ \mu_{S_{1}}^{\prime \; 2}  + (N - 1) \mu_{S_{2}}^{\prime \; 2}}  - [M_{\chi}^2]_{ii}}{(N - 1)} \qquad\qquad\qquad\qquad\qquad\qquad \text{ for $i \neq j$} \;. \nonumber 
\end{align} 
The eigenvalues of this matrix are
\begin{align}
m_{\xi_1}^{2}&= [M_{\chi}^2]_{ij} + (N-1)  \; [M_{\chi}^2]_{ii}    = 2 \lrb{ \mu_{S_{1}}^{\prime \; 2}  + (N - 1) \mu_{S_{2}}^{\prime \; 2}} \label{eq:masses_Gold_SN}\\
m_{\xi_i}^{2}&=[M_{\chi}^2]_{ii} - [M_{\chi}^2]_{ij}  \hspace{7cm}\text{ for } i=2,3...N \;.
\end{align}
The first ($m_{\xi_{1}}^{2}$) corresponds to the particle $\xi_{1}$, which is the PNGB ($m_{\xi_{1}}^2 \to 0$ as $\mu_{S_{1,2}}^{\prime \; 2} \to 0$), while the other particles ($\xi_{2,3,\dots, N}$)
are degenerate with mass $m_{\xi_{2}}=m_{\xi_{3}}=\dots=m_{\xi_{N}}$. As it turns out (in analogy to sec.~\ref{sec:U(1)*S_2}), the PNGB is the $S_{N}$-symmetric state
\begin{equation}
\xi_{1}=\frac{1}{\sqrt{N}} \sum\limits_{i=1}^{N} \chi_{i}\;, \label{eq:composition_Goldstone}
\end{equation}
where the others (not relevant to our discussion) can be found from orthonormality conditions.  We also note again that 
some hierarchy conditions should be imposed in order for the PNGB to be the DM particle.

\subsubsection*{The Cancellation of the Direct Detection Cross Section}
Again the ingredients that we need in order to show that the direct detection cross section vanishes,
are the inverse of the mass matrix for the real part of the scalars  and the interaction of a pair
of pseudo-Nambu-Goldstone particles with them (\ie $\xi_{1}^2 -h,\,s_{i}$). 

As usual the mass terms for the CP--even scalars can be written in a compact form as

\begin{equation}
\mathcal{L}_{\rm hs}=-\frac{1}{2} \Phi^{T} \, M^2_{\Phi} \, \Phi \; , 
\end{equation}
with $\Phi= \left(\begin{matrix} h\ s_{1} \ s_{2} \dots \end{matrix}\right)^{T}$ and 

\begin{align}
[M^2_{\Phi}]_{11} & = v^2 \lambda_{H} \; , \nonumber\\
[M^2_{\Phi}]_{1i} & = (M^2_{hs})_{i1} = v\,v_{S} \lrsb{ \lambda_{HS_{1}} + (N-1)\lambda_{HS_{2}} }  
& \text{ for } i>1 \; , \nonumber \\
[M^2_{\Phi}]_{ii} & =-v^2 (N - 1) \frac{\lambda_{HS_{2}}}{2} + \frac{N-1}{2} (\mu_{S_{2}}^2 + 
\mu_{S_{2}}^{\prime \, 2} ) + v_{S}^2 \Big[ 
\lambda_{S_{1}}                                                                  
 \label{eq:mass_matrix-SN}\\
&+ (N-1)c_{1} -\frac{1}{2}(N - 1) (N - 2) (c_{2} + 2 c_{2}^{\prime}) \nonumber\\
&- (N - 1) (N - 2) (N - 3) d \Big]  & \text{ for } 
i>1 \; , \nonumber \\
[M^2_{\Phi}]_{ij} & = v^2  \frac{\lambda_{HS_{2}}}{2} - \frac{1}{2}(\mu_{S_{2}}^2+\mu_{S_{2}}^{\prime \, 2}) + v_{S}^2 
\Big[  (\lambda_{S_{2}} + \lambda_{S_{1}}^{\prime} + 3 c_{1}) \nonumber\\
&+\frac{5}{2} (N - 2) (c_{2} +2 c_{2}^{\prime})+
3 (N - 2) (N - 3) d \Big]  & \text{ for } i,j>1 \;.\nonumber
\end{align}
The interaction term of the Lagrangian which is responsible for the $\xi_1 - N  $ elastic scattering is
\begin{align}
	\mathcal{L}_{\rm int}&= -\frac{1}{4N}\xi_{1}^2      
	\lrb{\begin{matrix} 
		Y_{\xi  h }
	     ,
		Y_{\xi  s }
		 ,
		 Y_{\xi s } 
		 , \dots\end{matrix}}   
	  \lrb{\begin{matrix} h \\ s_{1} \\ s_{2} \\ \vdots\end{matrix}} \; , 
	\label{eq:L_int_N}
\end{align}
with
\begin{align}
	Y_{\xi \, h}&=v N \lrsb{\lambda_{HS_{1}}  + (N-1) \lambda_{HS_{2}} } \\
	Y_{\xi \, s}&=v_{S} \Big\{   \lambda_{S_{1}} + (N-1) \lrsb{ \lambda_{S_{2} } + \lambda_{S_{1}}^{\prime} +4 c_{1}  + 2(N-2) \lrb{c_{2}+2c^{\prime} + (N-3)d}  } \Big\} \, .
\end{align}

Again, the propagator (\ie the inverse of the $s-h$ mass matrix) should be multiplied by a column vector $\sim \delta_{1i}$ (since only  $h$ interacts with SM fermions),
so the elements of the inverse of $M_{hs}^2$ relevant to the DM-nucleon interaction are 

\begin{align}
[M_{\Phi}^2]^{-1}_{11}&\sim    [M_{\Phi}^2]_{22} +(N-1)[M_{\Phi}^2]_{23}\;. \nonumber \\
[M_{\Phi}^2]^{-1}_{i1}&\sim - [M_{\Phi}^2]_{12}  \label{eq:Prop_SN}
\end{align}

As in sec.~\ref{sec:U(1)*S_2}, the Feynman diagram for the elastic PNGB-quark scattering is given fig.~\ref{fig:Diagram_ADD_N=2}, with an amplitude proportional to 

\begin{equation}
A_{\rm DD}\sim 	\lrb{\begin{matrix} 
	Y_{\xi  h }
	,
	Y_{\xi  s }
	,
	Y_{\xi s } 
	, \dots\end{matrix}}  \lrb{\begin{matrix} [M_{\Phi}^2]^{-1}_{11}  \\ [M_{\Phi}^2]^{-1}_{i1}  \\  [M_{\Phi}^2]^{-1}_{i1} \\ \vdots\end{matrix}} \;,
\end{equation}
which, from \eqs{eq:mass_matrix-SN, eq:L_int_N,eq:Prop_SN}, can be shown that vanishes.

\subsubsection*{A Note on the dark CP--invariance}

\begin{table}[h!]
\center
\begin{tabular}{| c ||| c |}
\hline
$N$ & \#phases \\
\hline\\[-0.45cm]
1 & 1\\
\hline\\[-0.45cm]
2 & 3\\
\hline\\[-0.45cm]
$\geq 3$ & $3+\frac{1}{2}N(N-1)$\\
\hline
\end{tabular}  
\caption{Number of phases for various values of $N$}
\label{phases}
\end{table}

In ref.~\cite{Gross:2017dan} it was argued that the $U(1)$ case is invariant under $S \to S^{\dagger}$, because there is one phase which can be absorbed by $S$. This  
natural symmetry of the model guarantees that the imaginary part of $S$ (the CP--odd scalar) always interact in pairs and as a result it is stable.  However, when  the scalar 
sector consists of a larger number of particles, it is not possible to absorb all phases to the scalars, as shown in Table~\ref{phases}. Therefore, in order to guarantee the 
stability of the  DM particle $\xi_{1}$, we have to  assume that all parameters are real on top of the $S_{N}$ symmetry.

\section{Conclusion and future direction} \label{sec:Conclusion}
\setcounter{equation}{0}
Inspired by an Abelian model which introduced a natural mechanism for the vanishing of the direct
detection cross section, we have expanded the discussion on the explanation of the smallness of the DM direct detection cross section.

The first case under study (sec.~\ref{sec:SU(2)}) was a softly broken $SU(2)$ global symmetry. 
In this, we assumed that there is a doublet scalar (singlet under the SM gauge symmetry),
which acquires a VEV. We showed that the resulting pseudo-Nambu-Goldstone bosons are all DM candidates,
due to a remaining discrete symmetry that  keeps them stable. We also showed that the DM--neucleon interaction  
vanishes. Then, we argued that this case can be generalized in a straightforward fashion to an $SU(N)$ symmetry, 
leading to the same result, \ie  vanishing of the DM--neucleon interaction.

Then in sec.~\ref{sec:U(1)*S_N}  we examined  the $U(1)\times S_{N}$ global symmetry, 
with $U(1)$ being softly  broken, where we extended the scalar sector by adding $N$ 
scalars, charged only under a global $U(1)$.
Assuming a dark CP--invariance, we calculated the form of the mass 
matrices and three-point interactions relevant to the pseudo-Nambu-Goldstone--nucleon interaction, which 
turned out to vanish.

A parameter space analysis of some simple cases (\eg $U(1)\times S_2$ or $SU(2)$), will help us identify
potential discovery channels at the LHC and astrophysical observations~\cite{Huitu:2018gbc,Alanne:2018zjm}.
Also, a calculation of 1-loop corrections will give us with precision the direct detection
cross section, which can further be used to probe (or even exclude) the models discussed in this work.
In addition, since the cases at hand should be treated as low-energy limits of complete models, an interesting direction would be to determine possible
completions. These, can induce (parametrically or energetically suppressed~\cite{Gross:2017dan,Alanne:2018zjm}) DM-nucleon interactions at the tree-level as well as 
decays of the PNGBs, allowing for a rich phenomenology, and connection of the DM problem with other open issues in particle physics (\eg lepton number violation and 
neutrino masses~\cite{Queiroz:2014yna}). 
Furthermore,  there are some cases that we did not consider (\ie the general irrep of the $SU(N)$ case), a study of other simple considerations (\eg $SU(2)$-triplet) 
can be insightful, and help us identify similar classes of models. 
However, since we were only interested in furthering the discussion on the suppression of the DM-nucleon interaction, with a focus on simple
realizations, we postpone these for a later project.

\section*{Acknowledgments}
DK is supported in part by the National Science Council (NCN) research grant No. 2015- 18-A-ST2-00748.
The author would like to thank Christian Gross, Alexandros Karam, Oleg Lebedev, and Kyriakos
Tamvakis, for their involvement at the early stages of this project.

\bibliography{refs}{}
\bibliographystyle{JHEP}                        

\end{document}